\documentclass[preprint,showpacs,amssymb,aps,prd]{revtex4}

\begin{document}

\def\diagram#1{{\normallineskip=8pt
       \normalbaselineskip=0pt \matrix{#1}}}

\def\diagramrightarrow#1#2{\smash{\mathop{\hbox to
.8in{\rightarrowfill}}
        \limits^{\scriptstyle #1}_{\scriptstyle #2}}}

\def\diagramleftarrow#1#2{\smash{\mathop{\hbox to .8in{\leftarrowfill}}
        \limits^{\scriptstyle #1}_{\scriptstyle #2}}}

\def\diagramdownarrow#1#2{\llap{$\scriptstyle #1$}\left\downarrow
    \vcenter to .6in{}\right.\rlap{$\scriptstyle #2$}}

\def\diagramuparrow#1#2{\llap{$\scriptstyle #1$}\left\uparrow
    \vcenter to .6in{}\right.\rlap{$\scriptstyle #2$}}

%
\title{SO(2,1) conformal anomaly:
\\
Beyond contact
interactions}

\author{Gino N. J. A\~{n}a\~{n}os,$^{1,2}$
 Horacio E. Camblong,$^{3}$
and
Carlos R. Ord\'{o}\~{n}ez$^{1,2}$}

\affiliation{
$^1$ Department of Physics, University of Houston, Houston,
TX 77204-5506
\\
$^2$
World Laboratory Center for Pan-American Collaboration in Science and
Technology,
\\
University of Houston Center, Houston, Texas 77204-5506
\\
$^3$
Department of Physics, University of San Francisco, San
Francisco, California 94117-1080}

\begin{abstract}
The existence of anomalous symmetry-breaking solutions of the 
SO(2,1)  commutator algebra is explicitly extended beyond 
the case of scale-invariant contact interactions. 
In particular, the 
failure of the conservation laws of the dilation and special conformal charges is 
displayed for the two-dimensional inverse square potential.
As a consequence, this anomaly appears to be a generic feature of conformal 
quantum mechanics and not merely an artifact of contact interactions.
Moreover, a renormalization procedure traces the emergence of this conformal 
anomaly to the ultraviolet sector of the theory, within which lies the apparent singularity.
\end{abstract}

\maketitle

\section{Introduction}
\label{sec:introduction}

The relevance of conformal quantum mechanics 
has been recognized for decades~\cite{alfaro_fubini_forlan:76} 
in the context of the scale-invariant Hamiltonian dynamics~\cite{jackiw:72}
of the inverse square potential,
which is characterized by an SO(2,1) commutator algebra.
A formally identical symmetry algebra was 
discovered for
the magnetic monopole~\cite{jackiw:80},
the magnetic vortex~\cite{jackiw:90},
and the two-dimensional
 contact interaction~\cite{jackiw:91-beg}.
Remarkably, this algebra
has
also been identified within the maximal 
``Schr\"odinger group'' of 
symmetries of nonrelativistic field theories~\cite{nr-SO(21)}
and related applications~\cite{jackiw:90b,bergman}.

Most importantly, 
the central role played by conformal quantum mechanics in theoretical physics
has been highlighted in recent years in a 
wide variety of problems. 
First,
insights into the physics of black holes have been directly gleaned from
the concept of near-horizon SO(2,1) conformal 
invariance~\cite{strominger:98,carlip:near_horizon,solodukhin:99},
as well as from its supersymmetric 
extensions~\cite{claus:98,deAz:99,michelson_strominger,papadopoulos:00,bellucci:02}.
This is in large part due to 
the remarkable connections provided by the
 AdS/CFT
correspondence~\cite{AdS/CFT}. 
In addition,
 the ubiquity of the Calogero model~\cite{calogero}, from black 
holes~\cite{calogero_black_holes}  to applications in 
condensed-matter physics~\cite{quantum_Hall,Haldane}, has led to
alternative applications of a formally identical algebra
 of conformal generators.
Finally, the use of field-theory renormalization techniques has
promoted novel methods for the treatment of singular 
interactions, including those within the conformal quantum mechanics 
class,
by means of Hamiltonian~\cite{jackiw:91-beg,cam:dtI,cam:dtII,gup:93,isp_letter,beane:00}
as well as path-integral techniques~\cite{pi_singular,green,pi_delta}.

The underlying property common to all the problems mentioned above
is the presence of a particular conformal symmetry 
after an appropriate {\em reduction\/}
framework is applied.
As such, this is a particular realization of conformal invariance
for an effective $(0+1)$-dimensional field.
It is the corresponding reduced
problem that is described within the conformal quantum
mechanics class, typically with an {\em effective Hamiltonian\/}
$
H
\equiv
 p^{2}/2M +V({\bf r})
$, or with many-body generalizations thereof.
In particular, in its reduced form,
a conformally invariant
interaction is characterized by an interaction potential $V({\bf r})$
that is a homogeneous function of degree $-2$.
This property alone implies that
these interactions satisfy a set of classical 
symmetries under time reparametrizations~\cite{cam:dtI,camblong:anom_delta}.
The associated
quantum-mechanical generators are
the Hamiltonian
$
H$,
the dilation operator
$
D
\equiv
tH
- \Lambda/2$,
in which  $\Lambda=
\left( {\bf p} \cdot {\bf r}
+  {\bf r} \cdot {\bf p}
\right)/2
$,
and the special conformal operator
$
 K
\equiv
2t D -
t^{2}H
+ 
M r^{2}/2
$;
these  generators yield
an
SO(2,1) Lie algebra~\cite{wyb:74} at the ``classical'' level
\begin{equation}
[D,H]_{\rm regular}
= - i \hbar H
 \;  ,
\;
\;  \;   \;   \;
[K,H]_{\rm regular}  = - 2 i \hbar D
\;  ,
\;
\;  \;   \;   \;
[D, K]_{\rm regular}  =  i \hbar K
\;  .
\label{eq:naive_commutators}
\end{equation}
Appropriate anomalous modifications of this ``regular'' 
algebra will  be discussed below,
when the theory is quantized.

The main purpose of this paper is to  explore
the quantum symmetry breaking of the algebra~(\ref{eq:naive_commutators})
for the inverse square potential.
The existence of this {\em conformal anomaly\/} was first recognized for
the two-dimensional contact interaction
by indirect methods in the seminal work of
Ref.~\cite{jackiw:91-beg} and was recently confirmed by a direct calculation
at the level of the commutator algebra 
in Refs.~\cite{camblong:anom_delta,esteve:anom_delta}.
Even though a draft
of the more general theory was developed in~\cite{camblong:anom_delta},
the proof of its actual realization for the 
all-important  inverse square potential  
is still lacking.
This is the problem to which we now turn our attention,
for the particular case of spatial dimensionality $d=2$.

\section{Ultraviolet Origin of the Anomaly for the Two-Dimensional Inverse Square
Potential}
\label{sec:ultraviolet_anomaly_ISP}

The 
inverse square potential
is of fundamental importance because of its applications to black 
holes~\cite{strominger:98,carlip:near_horizon,solodukhin:99,claus:98,deAz:99,michelson_strominger,papadopoulos:00,bellucci:02},
nuclear physics~\cite{beane:00,bedaque:99-three_body},
and molecular physics~\cite{molecular_dipole_anomaly}.
Even though the existence of this conformal anomaly had been
anticipated by other indirect arguments~\cite{molecular_dipole_anomaly}, in this work
we present the first conclusive direct
computation 
at the level of the commutator algebra~(\ref{eq:naive_commutators}). 
More precisely,
as the next step towards establishing
a more general framework, we show that the two-dimensional case of the
inverse square potential  
confirms the conclusions 
drawn in Ref.~\cite{camblong:anom_delta}.
The 
advantage of this particular dimensionality lies in the remarkable 
similarities that
the inverse square potential and 
the $\delta$-function interaction exhibit for $d=2$.
Not only is the dimensionality the same, but  both interactions are
characterized by 
a vanishing critical coupling, and the corresponding
expressions for the anomalous terms can be
considerably simplified. 

The fundamental quantity encoding the nature of the anomaly 
is~\cite{camblong:anom_delta}
\begin{equation}
{\mathcal A} ({\bf r})
\equiv
\frac{1}{i \hbar}
[D,H] +  H
=
\left[
\openone
+
\frac{1}{2}
\,
{\mathcal E}_{\bf r}
 \right]
V ({\bf r})
\;  ,
\label{eq:time_rate_of_dilation_op}
\end{equation}
where $\openone$
is the identity operator
and ${\mathcal E}_{\bf r} = {\bf r} \cdot 
{\bf \nabla} $
stands for the Eulerian derivative.
In particular,
the two-dimensional
form of Eq.~(\ref{eq:time_rate_of_dilation_op}) simplifies to
\begin{equation}
{\mathcal A} ({\bf r})=
\frac{1}{2}
\,
{\bf \nabla}
\!
\cdot
\!
\left\{
{\bf r} \,
 V ({\bf r})
\right\}
\;  .
\label{eq:time_rate_of_dilation_op_ddim}
\end{equation}
For the case of the two-dimensional inverse square potential, 
the 
Hamiltonian 
\begin{equation}
H= \frac{ p^{2}}{2M}  - \frac{g}{r^{2}}
\; 
\label{eq:2D_ISP_Hamiltonian_unregularized}
\end{equation}
is conformally invariant,
with $\lambda = 2M g/\hbar^{2}$ being the dimensionless form of the coupling constant.
Then, the formal two-dimensional identity
\begin{equation}
{\bf \nabla}
\!
\cdot
\!
\left[
\frac{
\hat{\bf r}}{r}
\right]
=
2 \pi \delta^{(2)} ({\bf r})
\;
\label{eq:Gauss_2D}
\end{equation}
implies that
\begin{equation}
{\mathcal A} ({\bf r})=
- g \,
\pi
\delta^{(2)} ({\bf r})
\;  ,
\label{eq:time_rate_of_dilation_op_ddim_2D_ISP}
\end{equation}
whose expectation value 
for a normalized state $\left| \Psi \right\rangle$ becomes
\begin{equation}
\frac{d}{dt}
\left\langle
D
\right\rangle_{\scriptstyle \!  \Psi}
=
\left\langle
{\mathcal A} ({\bf r})
\right\rangle_{\scriptstyle \!  \Psi}
=
- g \, \pi
\,
\int
d^{2} {\bf r}
\,
\delta^{(2)} ({\bf r})
\left| \Psi ({\bf r})
\right|^{2}
\;  .
\label{eq:time_rate_of_dilation_op_ddim-EV_2D_ISP}
\end{equation}

Equation~(\ref{eq:time_rate_of_dilation_op_ddim-EV_2D_ISP})
can be used to shed light on the nature of the possible conformal symmetry breaking.
Specifically,
 two important features are immediately apparent:

(i)  The correct evaluation of
Eq.~(\ref{eq:time_rate_of_dilation_op_ddim-EV_2D_ISP})
requires an appropriate regularization procedure, because of the
well-known 
vanishing or asymptotically free
value of $g$.
This behavior competes against 
the logarithmic singularity of the {\em renormalized\/}
ground-state wave function 
$\Psi ({\bf r})$ at the origin~\cite{cam:dtI,cam:dtII,isp_letter},
\begin{equation}
\Psi_{\! {\rm (gs)}}
 ({\bf r})
=
\frac{\kappa}{ \sqrt{\pi} } \, K_{0} (\kappa r)
 \; ,
\label{eq:2D_ISP_wf_normalized_renormalized}
\end{equation}
where $
\kappa=
\sqrt{2M
|E_{{\rm (gs)}}}|/\hbar$.
Consequently, 
Eq.~(\ref{eq:time_rate_of_dilation_op_ddim-EV_2D_ISP})
has to be regularized concurrently with 
other observables in the theory.

(ii) The existence of an anomaly
[nonvanishing value
of Eq.~(\ref{eq:time_rate_of_dilation_op_ddim-EV_2D_ISP})]
arises from the ``singularity'' at the origin, which is encoded in the 
$\delta$ function. The presence of this  generalized function
can be physically interpreted as 
representing the ``core'' of the interaction near the singular point,
according to Eq.~(\ref{eq:Gauss_2D}).
In short,
 the origin of this conformal 
anomaly can be conclusively traced to the apparent singularity at the origin, which lies within
the ultraviolet sector of the theory.

In the following sections, we will regularize 
the theory
using an ultraviolet  real-space regulator.

\section{Real-Space
Regularization of the Inverse Square Potential}
\label{sec:2D_ISP_circular_well_reg}

Real-space regularization
of the ultraviolet physics is implemented by an
appropriate 
modification of the interaction 
for $ r \alt a$.
This procedure amounts to the introduction of a 
regular potential $V^{(<)}({\bf r})  $ for 
$ r \alt a$, where it 
succinctly describes the short-distance physics.
Moreover, in order to maintain the intrinsic physics of the inverse square potential,
the core interaction $V^{(<)}({\bf r})  $ should
implement a continuous transition from
the long- to the short-distance physics~\cite{landau:77};
i.e., it should satisfy the continuity requirement $V^{(<)}(r=a) =-g/a^{2} $.  
The simplest and most convenient choice is afforded by
a finite square well 
$
V^{(<)}({\bf r})  
=
- g
\,
 \theta (a-r)/ a^{2}
$,
so that 
the unregularized Hamiltonian~(\ref{eq:2D_ISP_Hamiltonian_unregularized}) 
undergoes the replacement
\begin{equation}
H \rightarrow
H_{a}= \frac{ p^{2}}{2M}  
- 
\frac{g}{r^{2}} \,  \theta (r-a)
- 
\frac{g}{a^{2}}  \, \theta (a-r)
\; ,
\label{eq:2D_ISP_Hamiltonian_regularized}
\end{equation}
in which
 $\theta (z) $ stands for the Heaviside function.
Then, for a wave 
 function  $\Psi ({\bf r})= e^{im\phi} \, u_{|m|}(r)/\sqrt{r}$,
 the corresponding
 reduced radial Schr\"odinger equation 
is given by
\begin{equation}
\left\{
 \frac{d^2}{dr^2}
+ 
\left[ 
\frac{2M}{\hbar^{2}} \,   E +
\lambda  
\,
\frac{  \theta (a-r) }{ a^{2} }
 \right]
-
\frac{l^{2} 
- \lambda \, \theta (r-a)
-1/4 
}{r^2}
\right\}
u_{l} (r)
=
0
\;  ,
\label{eq:radial_Schr_2D_ISP_circular_well_reg}
\end{equation}
in which 
$l=|m|$, with $m$ being the usual 
quantum number.
A bound-state solution ($E<0$)
to Eq.~(\ref{eq:radial_Schr_2D_ISP_circular_well_reg})
can be written in terms of 
 Bessel functions~\cite{abr:72},
\begin{equation}
  R_{l}(r)
\equiv
\frac{ u_{l}(r) }{\sqrt{r}}
=\left\{\begin{array}{lr}
    \mbox{\boldmath\large  $\left\{  \right.$ } \! \! \!
J_{l} (\tilde k r)
\mbox{\boldmath\large  $,$ } \!   \!
N_{l} (\tilde k r)
\mbox{\boldmath\large  $\left.  \right\}$ } \! \!
\;  
& \textrm{for}
\;\; r<a  ,
\\
\mbox{\boldmath\large  $\left\{  \right.$ } \! \! \!
I_{i\Theta}(\kappa r)
\mbox{\boldmath\large  $,$ } \!   \!
 K_{i \Theta} (\kappa r) 
\mbox{\boldmath\large  $\left.  \right\}$ } \! \!
\;  
 & \textrm{for} \;\; r>a  , \\
  \end{array}
\right.
\;   \;   
\label{eq:Bessel_solution_2D_ISP_circular_well_reg}
\end{equation}
where  the effective coupling becomes
\begin{equation}
\Theta  \equiv \Theta_{l}  
= \sqrt{\lambda - l^{2}}
\; ,
\end{equation}
the energy parameters are
\begin{equation}
\tilde{k}^2
= 
\frac{2M}{\hbar^{2}} \,   
E
+\frac{\lambda}{ a^2} 
\;  
\label{eq:wave_number_2D_ISP_circular_well_reg}
\end{equation}
and
\begin{equation}
 \kappa^2
= - 
\frac{2M}{\hbar^{2}} \,   
E
\;  ,
\label{eq:kappa}
\end{equation}
and  the symbol
$\mbox{\boldmath\large  $\left\{  \right.$ } \! \! \!
\mbox{\boldmath\large  $,$ } \! \! \!
\mbox{\boldmath\large  $\left.  \right\}$ } \! \!
$
stands for linear combination.
In Eq.~(\ref{eq:Bessel_solution_2D_ISP_circular_well_reg})
the regular boundary conditions at the origin 
and at infinity lead to the particular selection 
\begin{equation}
 \Psi
  ({\bf r})=
e^{i m \phi}
\, \times
\left\{\begin{array}{lr}
    B_{l} \; 
J_{l}  (\tilde k r)  
\;  & \textrm{for} 
\;\;    r < a , \\ 
   A_{l} \; K_{i \Theta}  (\kappa r)  
\;   & \textrm{for}  \;\;  r  > a ,
\\
  \end{array}\right. 
\;  
\label{eq:wf_2D_ISP_circular_well_reg}
\end{equation}
where the relative values of $ A_{l} $ and $ B_{l}$  can be determined from the  continuity
condition
\begin{equation}
 B_{l} \,  J_{l}  ({\tilde k} a )
= A_{l} \, K_{i \Theta}  (\kappa a)
\; .
\label{eq:continuity_ISP_2D_circular_well}
\end{equation}
In addition,
the continuity of the logarithmic derivative  at  $r=a$
provides the equation for the energy eigenvalues
\begin{equation} 
\tilde k 
\,
\frac{J_l'(\tilde k  a)}{J_l(\tilde k  a)}
=
 \kappa 
\,
\frac{K_{i\Theta}'(\kappa  a)}{K_{i \Theta} (k a)}
\;  .
\label{eq:eigenvalue_2D_ISP_circular_well}
\end{equation}
Furthermore, the
 values of $A_{l}$ and $B_{l}$
  can be fixed from the normalization condition 
\begin{equation}
1
 = 
\int d^{2} {\bf  r} \, 
| 
\Psi
({\bf r})|^2  
 =
A_{l}^{2} \, 
2\pi
 \kappa^{-2} \, 
\left\{
{\mathcal K}_{i \Theta} (\kappa a) + 
\left( \frac{\kappa }{\tilde{k} } \right)^{2}
\,
\left[
\frac{
K_{i \Theta}  (\kappa a)  }{
J_{l}  ( \tilde{k} a ) }
 \right]^{2}
\,
{\mathcal J_{l}}( \tilde{k} a )  
\right\}
\; ,
\label{eq:normaliz_2D_ISP_circular_well_reg}
\end{equation}
where the functions
\begin{equation}
{\mathcal K_{i \Theta}} (\kappa a)
 = 
\int_{\kappa a}^{\infty} 
d z \, z
 \left[ K_{i \Theta} (z) \right]^{2} 
\; 
\label{eq:mathcal_K_integral_def}
\end{equation}
and
\begin{equation}
{\mathcal J_{l}} ( \tilde{k} a  )
 = 
\int_{0}^{  \tilde{k} a  } 
d z \, z
 \left[ J_{l} (z) \right]^{2} 
\; 
\label{eq:mathcal_J_integral_def}
\end{equation}
are conveniently defined.
Equations~(\ref{eq:normaliz_2D_ISP_circular_well_reg}),
(\ref{eq:mathcal_K_integral_def}),
and (\ref{eq:mathcal_J_integral_def})
will be further simplified when the theory is renormalized
in Sec.~\ref{sec:renormalization}.

\section{Calculation of the Conformal Anomaly}
\label{sec:circular_well}

We are now ready to start the computation of the regularized anomaly.
First, 
from Eqs.~(\ref{eq:time_rate_of_dilation_op_ddim}),
(\ref{eq:Gauss_2D}), and (\ref{eq:2D_ISP_Hamiltonian_regularized}),
the conformal anomaly manifests as 
the failure of the dilation operator to yield a zero time derivative;
explicitly,
 the regularized counterpart of Eq.~(\ref{eq:time_rate_of_dilation_op_ddim-EV_2D_ISP})
is obtained with the replacement
\begin{equation}
{\mathcal A} ({\bf r})
\rightarrow
{\mathcal A}_{a} ({\bf r})=
- g \,
\pi
\delta^{(2)} ({\bf r})
\,
  \theta (r-a) 
- \frac{g}{a^{2}} \,
\theta (a-r) 
\;  .
\label{eq:time_rate_of_dilation_op_ddim_2D_ISP_reg}
\end{equation}
Therefore,
the corresponding expectation value for a
 renormalized  and normalized state 
$\left| \Psi \right\rangle$ becomes
\begin{equation}
\frac{d}{dt}
\left\langle
D
\right\rangle_{\scriptstyle \!  \Psi}
=
\lim_{a \rightarrow 0}
\left[
\left\langle
{\mathcal A}_{a} ({\bf r})
\right\rangle_{\scriptstyle \!  \Psi_{a}}^{(<)}
+
\left\langle
{\mathcal A}_{a} ({\bf r})
\right\rangle_{\scriptstyle \!  \Psi_{a}}^{(>)}
\right]
\;  .
\label{eq:time_rate_of_dilation_op_ddim-EV_2D_ISP_reg}
\end{equation}
In Eq.~(\ref{eq:time_rate_of_dilation_op_ddim-EV_2D_ISP_reg}),
$\left| \Psi_{a} \right\rangle$ 
is the regularized counterpart of
$\left| \Psi \right\rangle$, 
as given in Eq.~(\ref{eq:wf_2D_ISP_circular_well_reg}).
Furthermore,
$
\left\langle
{\mathcal A}_{a} ({\bf r})
\right\rangle_{\scriptstyle \!  \Psi_{a}}^{(j)}
$
stands
for the contribution to the expectation value from the ultraviolet
region ($r<a$), for $j=<$; and from the region $r>a$, for $j=>$.
Remarkably,
Eq.~(\ref{eq:time_rate_of_dilation_op_ddim_2D_ISP_reg})
shows that 
\begin{equation}
\left\langle
{\mathcal A}_{a} ({\bf r})
\right\rangle_{\scriptstyle \!  \Psi_{a}}^{(>)}
= 0 
\; ,
\label{eq:anomaly_external_region}
\end{equation}
which
confirms that {\em only\/}
the singularity at the origin can 
be the source of the conformal anomaly.
As a consequence,
\begin{equation}
\frac{d}{dt}
\left\langle
D
\right\rangle_{\scriptstyle \!  \Psi}
 = 
\lim_{a \rightarrow 0}
\left\langle
{\mathcal A_{a}} ({\bf r})
\right\rangle_{\scriptstyle \!  \Psi_{a}}^{(<)}
=
- 2 \pi \,
\lim_{a \rightarrow 0}
\frac{g}{a^{2}} \, B_{l}^{2} \,
\int_{0}^{a} dr r \left[ J_{l} (\tilde{k} r) \right]^{2}
\;  .
\label{eq:anomaly_internal_region}
\end{equation}
This expression for the anomaly
can be most easily interpreted by rewriting it in the form
\begin{equation}
\frac{d}{dt}
\left\langle
D
\right\rangle_{\scriptstyle \!  \Psi}
 = 
E
\,
\lim_{a \rightarrow 0}
\left(
\left\{
\frac{\lambda (a)}{ (\tilde{k} a)^{2} }
\right\}
\,
\left\{
\frac{\pi A_{l}^{2} }{ \kappa^{2} } 
\right\}
\,
\left\{
\frac{ 2 
{\mathcal J_{l}} ( \tilde{k} a )}{ \left[ J_{l}(\tilde{k} a) \right]^{2} }
\,
\left[ K_{i \Theta} (\kappa a) \right]^{2}
\right\}
\right)
\;  ,
\label{eq:anomaly_2D_ISP_circular_well_calculation}
\end{equation}
as follows from 
Eqs.~(\ref{eq:kappa}), (\ref{eq:continuity_ISP_2D_circular_well}),
 and (\ref{eq:anomaly_internal_region}).
In Eq.~(\ref{eq:anomaly_2D_ISP_circular_well_calculation}),
$E$ is the finite renormalized 
value of the energy associated with $\left| \Psi  \right\rangle$,
and $\lambda (a)$ is the running coupling constant.
Correspondingly,
the anomalous time derivative of Eq.~(\ref{eq:anomaly_2D_ISP_circular_well_calculation}) 
is scaled with the bound-state energy $E$ of the state 
$\left| \Psi \right\rangle$.
Moreover, 
as we will show below, upon renormalization,
each one of the three additional factors enclosed in braces
is asymptotically equal to one
(with respect to the limit $a \rightarrow 0$).
As a result,
\begin{equation}
\frac{d}{dt}
\left\langle
D
\right\rangle_{\scriptstyle \!  \Psi}
= 
E
\;  ,
\label{eq:anomaly_2D_ISP_circular_well}
\end{equation}
which 
agrees with the
expected answer:
the right-hand side of Eq.~(\ref{eq:anomaly_2D_ISP_circular_well}) becomes the 
energy of the stationary normalized state~\cite{camblong:anom_delta}.

Finally, once the value of the
anomalous commutator $[D,H]$ has been identified,
the corresponding value of
the commutator $[K,H]$
is determined~\cite{camblong:anom_delta},
with 
\begin{equation}
\frac{d}{dt}
\left\langle
K
\right\rangle_{\scriptstyle \!  \Psi}
 =
2t
\, 
\frac{d}{dt}
\left\langle
D
\right\rangle_{\scriptstyle \!  \Psi}
= 2 t  E
\label{eq:D_K_conservation_law_failure}
\;  .
\end{equation}

\section{Renormalization}
\label{sec:renormalization}

The final required step is the renormalization of the system. 
This is implemented by finding the
behavior of the running coupling constant from  the consistency requirement that 
Eq.~(\ref{eq:eigenvalue_2D_ISP_circular_well})
admit a {\em finite\/} bound-state energy, when
  $a \rightarrow 0$.
From the small-argument expansion of the 
Macdonald function
\begin{equation}
K_{i\Theta} (z) 
  \stackrel{(z \rightarrow 0)}{=}
-
\sqrt{ \frac{\pi}{ \Theta 
\sinh  \left( \pi \Theta \right) } }
\,
\sin
\left(
\Theta
\left[
\ln \left( \frac{ z }{2} \right) 
+ \gamma
\right]
\right)
\,
\left[ 1 + O \left( z^{2} \right)  \right] 
\;  ,
\label{eq:macdonald_small_argument}
\end{equation}
Eq.~(\ref{eq:eigenvalue_2D_ISP_circular_well})
becomes
\begin{equation}
\Theta
\,
 \cot 
\left(
\Theta
\left[
\ln \left( \frac{ z }{2} \right) 
+ \gamma
\right]
\right)
\stackrel{(a \rightarrow 0)}{=}
\tilde{k} a
\,
\frac{ J'_{l} ( \tilde{k} a )}{  J_{l} ( \tilde{k} a )}
\,
\left[ 1 + O \left( [\kappa a]^{2} \right)  \right] 
\;   .
\label{eq:eigenvalue_2D_ISP_circular_well_limit_a=0}
\end{equation}

The renormalization condition consists in taking the limit $a \rightarrow 0$,
with a running coupling $\Theta (a)$ to be determined self-consistently
so that $\kappa$ remains fixed, thus guaranteeing a finite energy.
As argued in Refs.~\cite{gup:93} and \cite{isp_letter},
Eq.~(\ref{eq:eigenvalue_2D_ISP_circular_well_limit_a=0})
is ill defined unless  $\Theta (a)$ has the appropriate logarithmic running
\begin{equation}
\Theta  (a) \propto - [\ln (\kappa a)]^{-1} \stackrel{(a \rightarrow 0)}{\rightarrow} 0
\; .
\label{eq:asymptotic_freedom}
\end{equation}
 In other words, this  behavior drives the coupling $\lambda$ towards
its critical value, which is exactly zero for $d=2$; in particular,
when $l=0$,
$\lambda (a) = \Theta^{2} (a)  \stackrel{(a \rightarrow 0)}{\rightarrow} 0$.
Once this running behavior sets in, the only bound state that survives 
the renormalization process will occur for
$l=0$, because the other channels ($l \neq 0$) 
will be automatically placed in the weak-coupling regime, for which binding is 
suppressed~\cite{cam:dtII,isp_letter}.
In addition, this analysis shows that  binding will always occur for $d=2$, 
when the critical coupling is zero; this fact alone places the two-dimensional case in a unique 
position.
Moreover, the condition~(\ref{eq:eigenvalue_2D_ISP_circular_well_limit_a=0})
for the energy eigenvalues becomes
\begin{equation}
 \cot 
\left(
\Theta
\left[
\ln \left( \frac{ z }{2} \right) 
+ \gamma
\right]
\right)
\stackrel{(a \rightarrow 0)}{=}
-
\frac{\Theta}{2} \,
\left[ 1 + O (\Theta^{2}) \right]
\;   ,
\label{eq:eigenvalue_2D_ISP_circular_well_limit_a=0_approx}
\end{equation}
which logically enforces the limits $\cos \alpha  \stackrel{(a \rightarrow 0)}{\rightarrow} 0$
and $|\sin \alpha|  \stackrel{(a \rightarrow 0)}{\rightarrow} 1$, where $\alpha=
\Theta 
\left[
\ln (  z/2 ) 
+ \gamma
\right]
$.

We are now ready to prove
the fact that the three additional factors 
in Eq.~(\ref{eq:anomaly_2D_ISP_circular_well_calculation}) are asymptotically equal to one.
First, Eq.~(\ref{eq:wave_number_2D_ISP_circular_well_reg}) implies that
$
( \tilde{k} a )^{2} = \lambda + O( [\kappa a]^{2}) 
$,
in which $\lambda =\Theta^{2}$ is the leading logarithmic term
with respect to $a$, according to Eq.~(\ref{eq:asymptotic_freedom}); thus, 
 \begin{equation}
\frac{\lambda (a)}{ ( \tilde{k} a )^{2}} 
 \stackrel{(a \rightarrow 0)}{\rightarrow}
 1
\; ,
\end{equation}
for a finite energy level $E$.
The  second additional 
factor  in
Eq.~(\ref{eq:anomaly_2D_ISP_circular_well_calculation}) 
has a limiting value of one because
\begin{equation}
A_{0}
\stackrel{(a \rightarrow 0)}{=}
\frac{\kappa}{\sqrt{\pi}}
\left\{
1 + 
o (\kappa a)
\right\}
\;  .
\label{eq:Acoeff_2D_ISP_circular_well}
\end{equation}
This can be deduced from Eqs.~(\ref{eq:normaliz_2D_ISP_circular_well_reg}),
(\ref{eq:mathcal_K_integral_def}),
and (\ref{eq:mathcal_J_integral_def}),
for $l=0$, $\kappa a \ll 1$, and $\Theta \sim \tilde{k} a \ll 1$, 
which collectively imply that
$2 \, {\mathcal K_{i \Theta}} (\kappa a) = 1 + O(\Theta^{2})$
and
$ 2 \, {\mathcal J_{l=0}}  (\tilde{k} a )
= \Theta^{2}
\left[ 1 + O(\Theta^{2}) \right]$.
Finally, the third  additional factor 
in Eq.~(\ref{eq:anomaly_2D_ISP_circular_well_calculation}) 
becomes
\begin{equation}
2 \,
\frac{\mathcal J_{l} 
( \tilde{k} a )}{ \left[ J_{l}(\tilde{k} a) \right]^{2} }
\,
\left[ K_{i \Theta} (\kappa a) \right]^{2}
\stackrel{(a \rightarrow 0)}{=}
2 \,
 \left( \frac{\Theta^{2} }{2} \right)  \,
\left[- \frac{\sin \alpha}{\Theta} \right] ^{2} 
\,
\left[ 1 + O(\Theta^{2}) \right]
\stackrel{(a \rightarrow 0)}{=}
1 + O(\Theta^{2}) 
\; .
\end{equation}

In closing, this renormalization
procedure, based on the modification of the ultraviolet  behavior,
shows that the  ground state wave function
reduces to Eq.~(\ref{eq:2D_ISP_wf_normalized_renormalized})
in the limit $a \rightarrow 0$.
However, as discussed in Sec.~\ref{sec:ultraviolet_anomaly_ISP},
for the computation of the anomaly, this limit can only be taken 
as the last step, after all expressions have been properly regularized.
In this paper we have shown that
the ensuing procedure
is implemented at the level of Eq.~(\ref{eq:anomaly_2D_ISP_circular_well_calculation})
and yields the anticipated answer, Eq.~(\ref{eq:anomaly_2D_ISP_circular_well}).

\section{Conclusions}
\label{sec:conclusions}

In conclusion, we have shown  the 
existence of a conformal anomaly of the SO(2,1) algebra
associated with the dynamics of the two-dimensional inverse square potential.
The corresponding violations of the conservation laws of the 
dilation and special conformal charges
follow patterns very similar to those encountered earlier for contact interactions.
In particular, this work 
is closely related to 
the conformal interactions of maximal physical relevance,
involved in applications from molecular physics to the physics of black 
holes.
Consequently,  this analysis leads to
 new insights into the emergence of
anomalies within the framework of conformal quantum mechanics.
 Finally, these ideas can be generalized beyond the two-dimensional case and
for a more general modification of the ultraviolet physics;
additional details are in progress and will be reported elsewhere.

\acknowledgments{This research was supported in part by
an Advanced Research Grant from the Texas
Higher Education Coordinating Board 
and by the University of San Francisco Faculty Development Fund.
One of us (G.N.J.A.)
gratefully acknowledges the generous support from the World Laboratory.
}

\end{document}